\documentclass[screen,sigconf]{acmart}
\AtBeginDocument{%
  }

\usepackage{newfloat}

\floatstyle{plain}
\newfloat{diagram}{thd}{lod}
\floatname{diagram}{Diagram}

\usepackage[section]{minted}

\usepackage{calc}

\usepackage{multirow}

\usepackage{xspace}

\usepackage{suffix}

\usepackage[framemethod=TikZ]{mdframed}

\usepackage{xpatch}

\usepackage{tikz}
\usetikzlibrary{arrows.meta}

\usepackage[nameinlink,noabbrev]{cleveref}

\newlength{\FramedXMargin}
\newlength{\FramedYMargin}
\newlength{\FramedSkipAbove}
\newlength{\FramedSkipBelow}

\newsavebox{\FramedBox}

\newenvironment{Framed}{%
  \begin{lrbox}{\FramedBox}
  \begin{minipage}{\dimexpr\linewidth-2\fboxrule-2\fboxsep} 
  \vspace{\FramedYMargin}%
  \centering%
  \begin{minipage}{\dimexpr\linewidth-2\FramedXMargin} 
}{%
  \end{minipage}%
  \vspace{\FramedYMargin}
  \end{minipage}%
  \end{lrbox}%
  \vspace{\FramedSkipAbove}
  \noindent\fbox{\usebox{\FramedBox}}
  \vspace{\FramedSkipBelow}
}

\newcounter{RQ}

\newcounter{SubRQ}
\renewcommand{\theSubRQ}{\arabic{SubRQ}}

\newenvironment{MainResearchQuestion}{%
  \refstepcounter{RQ}%
  \begin{Framed}\label{rq:main}
  \textbf{RQ:}%
}{%
  \end{Framed}
}

\newenvironment{SubResearchQuestion}{%
  \refstepcounter{SubRQ}%
  \begin{Framed}\label{srq:\theSubRQ}%
  \textbf{sRQ\theSubRQ:}%
}{%
  \end{Framed}%
}

\newenvironment{RevisitMainResearchQuestion}{%
  \begin{Framed}
  \textbf{RQ:}%
}{%
  \end{Framed}%
}

\newenvironment{RevisitSubResearchQuestion}[1]{%
  \begin{Framed}
  \textbf{sRQ#1:}%
}{%
  \end{Framed}%
}

\crefformat{RQ}{#2RQ#3}
\Crefformat{RQ}{#2RQ#3}
\crefformat{SubRQ}{#2sRQ#1#3}
\Crefformat{SubRQ}{#2sRQ#1#3}

\crefrangeformat{SubRQ}{#3sRQ#1#4 to #5sRQ#2#6}
\Crefrangeformat{SubRQ}{#3sRQ#1#4 to #5sRQ#2#6}

\emergencystretch=\maxdimen
\newcommand{\DW}{\textsc{Digital Wallet}\xspace}
\newcommand{\Wallet}[1]{\textsc{#1 Wallet}\xspace}
\newcommand{\Transaction}[1]{\textsc{#1}\xspace}

\newcommand{\Entity}[1]{\texttt{#1} entity\xspace}
\newcommand{\EntityR}[1]{\texttt{#1}\xspace}

\newcommand{\Entities}[1]{\texttt{#1} entities\xspace}
\newcommand{\EntitiesR}[1]{\texttt{#1}\xspace}

\newcommand{\Service}[1]{\texttt{#1} service\xspace}

\newcommand{\Variable}[1]{\texttt{#1}\xspace}

\newcommand{\Status}[1]{\textit{#1} status\xspace}
\newcommand{\StatusR}[1]{\textit{#1}\xspace}

\newcommand{\TypeR}[1]{\textit{#1}\xspace}

\newcommand{\Request}[1]{\textit{#1}\xspace}

\newcommand{\Object}[1]{\texttt{#1}\xspace}
\newcommand{\Class}[1]{\texttt{#1}\xspace}
\newcommand{\Method}[1]{\texttt{#1}\xspace}

\newcommand{\sRQ}[1]{\texorpdfstring{\cref{srq:#1}}{sRQ#1}}

\DeclareFloatingEnvironment[
  fileext=lol,
  name=Listing
]{code}

\crefname{code}{listing}{listings}
\Crefname{code}{Listing}{Listings}

\newmintedfile[InputKotlin]{kotlin}{
  frame=topline,
  framesep=2mm,
  baselinestretch=1.15,
  fontsize=\scriptsize,
  linenos,
  numbersep=0.5em,
  frame=lines,
  framesep=0.5em,
}

\newmintedfile[InputScala]{scala}{
  frame=topline,
  framesep=2mm,
  baselinestretch=1.15,
  fontsize=\scriptsize,
  linenos,
  numbersep=0.5em,
  frame=lines,
  framesep=0.5em,
}

\setcopyright{acmlicensed}
\copyrightyear{2025}
\acmYear{2025}
\acmDOI{XXXXXXX.XXXXXXX}

\acmConference[SBQS25]{Brazilian Symposium on Software Quality}{Nov 04--07,
2025}{São José dos Campos, SP}
\setcopyright{none}
\renewcommand\footnotetextcopyrightpermission[1]{}

\renewcommand\abstractname{ABSTRACT}
\renewcommand\keywordsname{KEYWORDS}
\renewcommand\refname{REFERENCES}

\begin{document}

\title[
  Functional vs. Object-Oriented
  on the Architectural Characteristics of Systems
]{
  Functional vs. Object-Oriented:\texorpdfstring{\\}{ }
  Comparing How Programming Paradigms Affect the\texorpdfstring{\\}{ }
  Architectural Characteristics of Systems
}

\author{Briza Mel Dias}
\orcid{0009-0004-6211-3307}
\affiliation{%
  \institution{University of São Paulo}
  \city{São Paulo}
  \country{Brazil}
}
\email{brizamel.dias@alumni.usp.br}

\author{Renato Cordeiro Ferreira}
\orcid{0000-0001-7296-7091}
\affiliation{%
  \institution{Jheronimus Academy of Data Science}
  \city{'s-Hertogenbosch}
  \country{The Netherlands}
}
\affiliation{%
  \institution{University of São Paulo}
  \city{São Paulo}
  \country{Brazil} \\
}
\email{renatocf@ime.usp.br}

\author{Alfredo Goldman}
\orcid{0000-0001-5746-4154}
\affiliation{%
  \institution{University of São Paulo}
  \city{São Paulo}
  \country{Brazil}
}
\email{gold@ime.usp.br}

\renewcommand{\shortauthors}{Dias et al.}

\begin{abstract}
  This study compares the impact of adopting object-oriented programming
  (OOP) or functional programming (FP) on the architectural characteristics
  of software systems.
  For that, it examines the design and implementation of a \DW system
  developed in Kotlin (for OOP) and Scala (for FP).
  The comparison is made through a mixed-method approach.
  The self-ethnographic qualitative analysis provides a
  side-by-side comparison of both implementations,
  revealing the perspective of those writing such code.
  The survey-based quantitative analysis gathers feedback
  from developers with diverse backgrounds,
  showing their impressions of those reading this code.
  Hopefully, these results may be useful for developers
  seeking to decide which paradigm is best suited for their
  next project.
\end{abstract}



\maketitle

\section{Introduction}
\label{sec:introduction}

After decades of dominance of Object-Oriented Programming (OOP), the Functional
Programming (FP) paradigm has gained significant attention in the software
industry~\citep{Site:Github}. This shift is a response to growing complexities
in software systems, the demand for better scalability, and the need for more
predictable, robust, and maintainable codebases.

According to the IEEE Spectrum magazine in 2022 \cite{Site:IEEESpectrum},
one of the clearest indications of this shift is the increasing incorporation
of functional features into different mainstream programming languages.
Evaluating how FP compares to OOP is essential to understanding the
future of software development.




The main research question of this paper can be summarized as:
\begin{MainResearchQuestion}
  How do the functional and object-oriented paradigms impact different
  architectural characteristics of a system?
\end{MainResearchQuestion}

According to \citeauthor{Richards:FundamentalsSoftwareArchitecture:2020}
\cite{Richards:FundamentalsSoftwareArchitecture:2020}, an \emph{architectural
characteristic} describes a concern that follows three criteria:
\begin{enumerate}
  \item it specifies a non-domain design consideration,
  \item it influences structural aspects of the design, and
  \item it is critical or important to the application's success.
\end{enumerate}



This research focuses on six architectural characteristics,
summarized in \cref{tab:architectural_characteristics}.
They encompass concerns that highlight important differences between the
paradigms, making it easier to compare their strengths and limitations.

\begin{table}[h]
\centering
\caption{Evaluation Criteria for Qualitative Analysis}
\vspace{-0.6\baselineskip}
\begin{tabular}{|p{2.5cm}|p{5cm}|}
  \hline
  \textbf{Architectural Characteristic}
  & \multirow{2}{*}{ \textbf{From the code, how easy it is to...} } \\ \hline
  Extensibility      & ...add new behavior                   \\ \hline
  Reusability        & ...apply it in a new use case         \\ \hline
  Error Handling     & ...understand errors handled by it    \\ \hline
  Error Propagation  & ...understand errors propagated by it \\ \hline
  Testability        & ...test its intended behavior         \\ \hline
  Readability        & ...understand its intended behavior   \\ \hline
\end{tabular}
\label{tab:architectural_characteristics}
\end{table}

Due to space constraints, this paper will address only four of the six
architectural characteristics from \cref{tab:architectural_characteristics},
whose analyses brought the most interesting results. For a full discussion,
please refer to the bachelor thesis that originated this paper%
~\cite{Dias2024FunctionalSystems}.



\begin{SubResearchQuestion}
How do the functional and object-oriented paradigms impact
the extensibility of a system? 
\end{SubResearchQuestion}

\Cref{srq:1} aims to explore how each paradigm influences the ability
to add new components to the system. OOP typically relies on object 
composition and class inheritance, while FP uses high-order functions
and generic data structures.

\begin{SubResearchQuestion}
How do the functional and object-oriented paradigms impact
the reusability of a system? 
\end{SubResearchQuestion}

\Cref{srq:2} aims to explore how each paradigm influences the ability
to reuse components in the system. OOP typically relies on design
patterns that take advantage of polymorphism, while FP combines
list-based immutable data types with closures and recursion.

\begin{SubResearchQuestion}
How do the functional and object-oriented paradigms impact
error handling and propagation of a system? 
\end{SubResearchQuestion}

\Cref{srq:3} aims to explore how each paradigm embodies distinct
philosophies in error management. OOP typically uses exceptions,
while FP applies monadic types.




\section{Programming Paradigms}
\label{sec:paradigms}

Understanding the principles of OOP and FP is essential for a meaningful
comparison between them. Their distinct philosophies, strengths, and
limitations significantly influence how software systems approach
architectural characteristics.

\subsection{Object-Oriented Programming}
\label{subsec:oop}

According to \citeauthor{Martin:CleanArchitecture:2017}%
~\cite{Martin:CleanArchitecture:2017}, any OOP language must support
these three characteristics: \emph{encapsulation},
\emph{inheritance}, and \emph{polymorphism}.

\subsubsection*{Encapsulation}\label{oop:encapsulation}
It ensures that a clear boundary is established around a cohesive set of data
and behaviors~\cite{Martin:CleanArchitecture:2017}. Within this
boundary, data is hidden and protected from direct external access,
while only specific functions are exposed to interact with the data.
This approach is commonly seen in object-oriented programming through
private data members and public methods in classes.

\subsubsection*{Inheritance}\label{oop:inheritance}
It is the process of reusing and extending a group of variables
and functions defined in one scope within another%
~\cite{Martin:CleanArchitecture:2017}.
In OOP, this often means that a new class (called a subclass) can inherit
properties and behaviors from an existing class (called a superclass).
This allows the new class to reuse, override, or expand existing
functionality without redefining it entirely.

\subsubsection*{Polymorphism}\label{oop:polymorphism}
It is the ability of a method to behave differently based on the
context in which it is used~\cite{Martin:CleanArchitecture:2017}.
Polymorphism unblocks \emph{dependency inversion}
by allowing high-level modules to depend on abstractions rather than concrete
implementations. It decouples implementation details from the modules that
depend on them through interfaces, allowing dependencies to be reversed.

\subsection{Functional Programming}\label{subsec:fp}

According to \citeauthor{Pilquist:FunctionalProgrammingScala:2023}%
~\cite{Pilquist:FunctionalProgrammingScala:2023},
functional programming is based on the premise of writing programs
using only \emph{pure functions}, which leads to other patterns
like \emph{high-order functions} and \emph{monadic types}.

\subsubsection*{Pure Functions}\label{fp:pure_functions}
They are functions without \emph{side effects}%
~\cite{Martin:CleanArchitecture:2017}.
A function has side effects if it does something other than returning a result,
such as modifying a variable, throwing an exception, or printing to the console.

\subsubsection*{High-Order Functions}\label{fp:hof}
In FP, functions are values%
~\cite{Pilquist:FunctionalProgrammingScala:2023}.
They can be assigned to variables, stored in data structures, and passed
as arguments to functions. A function that accepts other functions as
arguments is called a higher-order function (HOF).

\subsubsection*{Monadic Types}\label{fp:monadic_types}
Most FP languages support \emph{pattern matching}%
~\cite{Pilquist:FunctionalProgrammingScala:2023}
as a way to execute functions depending on the value passed to them.
Constructs such as \texttt{Option} and \texttt{Either} have the properties
of Monads~\cite{Pilquist:FunctionalProgrammingScala:2023}, types that 
leverage pattern matching to provide clean composition and error handling.
\section{Related Literature}
\label{sec:related_literature}

Few works in the literature focus on the same research questions
proposed in \cref{sec:introduction}. A search with the keywords
\href{https://scholar.google.com/scholar?q=functional+vs+object+oriented+programming}%
{``Functional versus Object-Oriented Programming''} on on Google Scholar
results in papers that propose to mix the paradigms in the context of a
single programming language or application.

In the late 1990s, \citeauthor{Harrison1996ComparingPrograms}%
~\cite{Harrison1996ComparingPrograms,Samaraweera1998EvaluationParadigms}
compared the code of $12$ programs produced in SML (FP) and C++ (OOP),
measuring their difference on development and code metrics.
In 2016, \citeauthor{Alic2016ComparativeProgramming}%
~\cite{Alic2016ComparativeProgramming}
compared algorithms in Java and C\# (OOP) versus Haskell and F\# (FP),
focusing on lines of code and efficiency.
In 2025, \citeauthor{Khan2025AArchitecture}~\cite{Khan2025AArchitecture}
compared the impact of choosing OOP, FP or Procedural Programming
for implementing microservices architectures%
~\cite{Newman:BuildingMicroservices:2021},
analyzing their impact on application design.

The current gap in the literature is an opportunity to explore the
differences between paradigms once again, using modern programming
languages and focusing on both code and design.

\section{Methodology}\label{sec:methodology}

This section details the research methodology employed in this project,
whose goal is to answer \cref{rq:main}.

\subsection{Proof of Concept}\label{subsec:methodology_proof_of_concept}

The first phase of the research consists of the specification and
development of a proof-of-concept system that can be used as an object
of study. This provides a detailed example for comparing the
object-oriented and functional paradigms.

A \DW was chosen as the proof-of-concept system for this research.
It contains many functionalities, such as
  wallet management,
  transaction management, and
  calculation of balances.
These operations demand intricate data handling to keep consistency,
whose implementation can be approached differently within
the object-oriented and functional paradigms.

To compare the paradigms, the same \DW system was developed \emph{twice},
following the good practices of each paradigm.

\subsubsection*{Object-Oriented PoC}\label{poc:kotlin}
The object-oriented version of the \DW system was based on the
\href{https://kotlinlang.org/}{Kotlin} programming language,
which supports the OOP principles from \cref{subsec:oop}.
The developer can define classes in a straightforward manner,
enabling the structuring of data and behavior through objects.

\subsubsection*{Functional PoC}\label{poc:scala}
The functional version of the \DW system was based on the
\href{https://www.scala-lang.org/}{Scala} programming language,
which supports classic FP principles from \cref{subsec:fp}.
The developer can focus on immutability and pure functional constructs,
aligning with the core principles of functional programming.

\medskip
\Cref{sec:requirements} describes the set of functional requirements that
guided the development of both versions of the \DW system.
\Cref{sec:system_entities,sec:system_services} describe the main business
logic that was implemented in both proofs of concept.
The code developed during the research is available in the paper's
\href{https://doi.org/10.5281/zenodo.16618132}{reproduction package}.

\subsection{Analysis}
\label{subsec:methodology_analysis}

The second phase of the research consists of an analysis of the paradigms
using a mixed-method approach. By triangulating benefits and drawbacks of
each paradigm over the architectural characteristics, it is possible to
provide an answer to \cref{rq:main}.

The analysis will be based on the implementation of the \DW system.
In particular, the focus is on code snippets that perform the same task
from both proofs of concept.

To provide a good perspective over the paradigms, this research proposes
to combine two research techniques:


\subsubsection*{Self-Ethnographic Qualitative Analysis}
It represents the perspective of \emph{writing code} in the different
paradigms. In particular, the self-ethnographic methodology means that it
presents the assessments of the authors of the paper, who developed the
proofs of concept.
These results are further explored in \cref{sec:qualitative_analysis}.

\subsubsection*{Survey-Based Quantitative Analysis}
It represents the perspective of \emph{reading code} in the different
paradigms. In particular, the survey-based methodology means that it
summarizes the assessments of different developers with varying levels
of experience with the paradigms.
These results are further explored in \cref{sec:quantitative_analysis}.

\medskip
After presenting these results, \cref{sec:discussion} combines these
perspectives to answer the research subquestions.


\section{Requirements}\label{sec:requirements}

The \DW system is the object of study used in this work to establish the
comparison between the object-oriented and the functional programming paradigms.

Although it models a real-world problem, the \DW system was not developed for
production but rather was designed as a proof of concept whose functionalities
highlight the differences between the paradigms. Therefore, the \DW system
ignores some important non-functional requirements for this type of
system, such as security and auditability.
  
The functional requirements outlined for the \DW system are listed below.

\subsection{Wallet Management}
\label{subsec:requirement_wallet_management}
      
The \DW system should provide each customer with three different types of
wallets:

\subsubsection*{\Wallet{Real Money}}
It represents a real-world checking account owned by the customer.
This wallet is the entry point of the \DW system, where the customer can deposit
from or withdraw to an external bank account. It is modeled to be a temporary
allocation of the customer’s funds until they are moved to another wallet.

\subsubsection*{\Wallet{Investment}}
It represents a portfolio of investment options where customers can allocate
their funds. Funds are invested by transferring funds from the \Wallet{Real
Money} to the \Wallet{Investment} and are liquidated the other way around.
Investments and liquidations can be initiated at any time but will only settle
on the next business day.

\subsubsection*{\Wallet{Emergency Funds}}
It represents deposit insurance where customers can allocate funds without
taking the risks of the \Wallet{Investment}. Funds are deposited into and
withdrawn from the \Wallet{Emergency Funds} through instant transfers to and
from the \Wallet{Real Money}.
 
\subsection{Transaction Lifecycle Management}
\label{subsec:requirement_transaction_lifecycle_management}

\begin{figure}[b]
  \centering
  \includegraphics[width=0.98\linewidth]{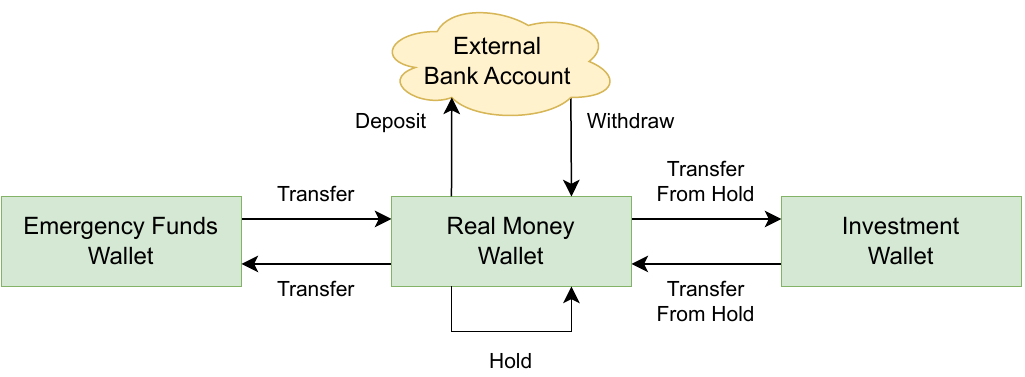}
  \Description{}
  \caption{
    Transaction Lifecycle.
  }%
  \label{fig:transactions_and_wallets}
\end{figure}

The \DW system should support the creation, validation, and execution of
transactions between wallets. Transactions represent the flow of money
inside the \DW system, which might succeed or fail.

To provide customers with all the functionalities, the \DW system needs
to support the following transactions:

\begin{itemize}
  \item \Transaction{Deposit},
        which represents a deposit to the~\Wallet{Real Money},
        modeling the entry of funds into the \DW system.
  \item \Transaction{Withdrawal},
        which represents a withdrawal from the \Wallet{Real Money},
        modeling the exit of funds from the \DW system.
  \item \Transaction{Transfer},
        which represents an instant transfer to or from the \Wallet{Real Money}.
        While the \DW system is not required to handle transfers directly
        with real-world bank accounts, it should integrate with a third-party
        API to enable this functionality.
  \item \Transaction{Hold},
        which represents an amount of money reserved for future use.
        This transaction ensures funds until an application settles.
  \item \Transaction{Transfer\;From\;Hold},
        which represents an instant release of an amount of money
        previously reserved in a~\Transaction{Hold}. This transaction
        settles an application, using the available funds.
\end{itemize}

\Cref{fig:transactions_and_wallets} summarizes the compatibility between 
transactions and different wallets.
Every transaction must be validated before execution.
\Transaction{Deposit} and \Transaction{Withdrawal} transactions
are restricted to the \Wallet{Real Money}.
\Transaction{Transfer} transactions are allowed only between
\Wallet{Real Money} and \Wallet{Emergency Funds}.
\Transaction{Hold} transactions can be placed only on \Wallet{Real Money}
and~\Wallet{Investment}, whereas \Transaction{Transfer\;From\;Hold}
transactions are limited to being executed between them.
Finally, all transactions -- except for \textsc{Deposit} -- require a
balance check to ensure sufficient funds are available to complete the
operation.

The \DW system must manage transaction failures as part of the
transaction lifecycle. Transactions failing validation should be
marked as \emph{permanently failed}, while those that fail during
execution should be eligible for \emph{retry}. In general, validation
should capture expected, unrecoverable errors, such as insufficient funds. Execution errors, though unlikely after successful validation, may still
occur due to internal issues or third-party API instabilities. In such cases,
the system should allow to retry the transaction.

\subsection{Investment Customization}
\label{subsec:requirement_investment_customization}

      
The \DW system should allow customizing investments and liquidations.
The participation should be a percentage of the total invested under
an \Wallet{Investment}, and the customer should have real-time control
over it. This requirement includes the capacity of setting the percentage
to zero for undesired investments or even selecting a single investment.

\subsection{Minor Requirements}
\label{subsec:requirement_minor}

\subsubsection*{Financial Tracking}
The \DW system must track all financial transactions to maintain accurate
account balances and ensure data integrity.

\subsubsection*{Investment and Liquidation Settlement}
The \DW system must settle all investment and liquidation requests initiated
on the previous business day.

\subsubsection*{Batch Processing of Transactions}
The \DW system must support batch processing of transactions with atomicity.
In any given batch, all transactions must either complete successfully or,
if any transaction fails, the entire batch must be rolled back.
This ensures data integrity and prevents partial updates that could disrupt
the financial record-keeping system.

\section{\DW System: Entities}
\label{sec:system_entities}




This section presents entities and value objects%
~\cite{Evans:DomainDrivenDesign:2003} representing the core
business concepts of the \DW system.

\subsection{Wallet}
\label{subsec:entity_wallet}

The \Entity{Wallet} models an account where customer funds are held and
managed. It functions as the hub for all financial transactions, allowing
money to be deposited or withdrawn to other destinations.
The \DW system provides three types of wallets, as specified in
\cref{subsec:requirement_wallet_management}:~
\EntityR{Real\;Money\;Wallet}, \EntityR{Investment\;Wallet},
and \EntityR{Emergency\;Fund\;Wallet}.

While \Entities{Transaction} dictate the flow of funds between
\Entities{Wallets}, the latter are unaware of the former's existence.
Their role is to model and hold customer funds within the system.
They are not responsible for determining the validity of any incoming
or outgoing \EntityR{Transaction}. This separation of concerns ensures
that \Entities{Wallet} focus on balance management without handling
\EntityR{Transaction} logic.

\subsection{Subwallet}
\label{subsec:entity_subwallet}

The \textsc{Subwallet} concept is introduced to represent the distribution of
funds within the same wallet. 

The \EntityR{Real\;Money\;Wallet} has only one \Entity{Subwallet}:
the \EntityR{Real\;Money\;Subwallet}.
Similarly, the \Entity{Emergency\;Funds\;Wallet} has only one \Entity{Subwallet}:
the \EntityR{Emergency Funds Subwallet}.
In practice, having a single \EntityR{Subwallet} means that,
for record-keeping, funds in the \EntityR{Wallet} are not divided in any way.

\begin{figure}[th!]
  \centering
  \includegraphics[width=0.70\linewidth]{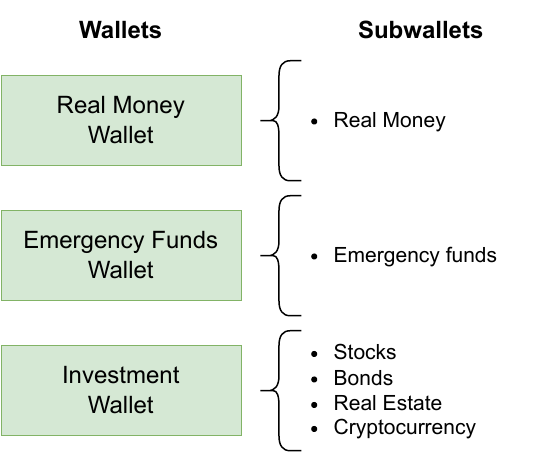}
  \Description{
    \EntitiesR{Wallets} and corresponding \EntitiesR{Subwallets}.
  }
  \caption{
    \EntitiesR{Wallets} and corresponding \EntitiesR{Subwallets}.
  }
  \label{fig:wallet_and_subwallets}
\end{figure}

The relevance of the \Entity{Subwallet} becomes evident in the
\EntityR{Investment Wallet}, where funds are allocated across various
investment options, as illustrated by \cref{fig:wallet_and_subwallets}.
Introducing a new investment option involves composing new
\Entities{Subwallet} within the \EntityR{Investment Wallet}.

\subsection{Transaction}
\label{subsec:entity_transaction}

A \Entity{Transaction} represents an attempt to change the state of a
\EntityR{Wallet}. Each transaction specifies an \Variable{amount},
as well as \Variable{originator} and \Variable{beneficiary}, which can
either be a \EntityR{Wallet} instance (and its designated \EntityR{Subwallet})
or an external bank account.
Each \EntityR{Transaction} also has a \Variable{status} representing its
current phase in processing. Possible statuses include \StatusR{Processing}, \StatusR{Failed}, \StatusR{Transient Error}, and \StatusR{Completed}. 

\subsubsection*{Transaction Types} 
A \EntityR{Transaction} can be classified as a
  \TypeR{Deposit},
  \TypeR{Withdrawal},
  \TypeR{Hold},
  \TypeR{Transfer}, or
  \TypeR{Transfer~From~Hold}.
Each transaction type represents a standard operation that modifies
the state of a \EntityR{Wallet} as specified in
\cref{subsec:requirement_transaction_lifecycle_management}.

\subsubsection*{Transaction validation}
A \EntityR{Transaction} must be validated to ensure they adhere
to the internal rules of the \DW system. For example, a
\TypeR{Deposit} and \TypeR{Withdrawal} can only be initiated on a
\EntityR{Real Money Wallet}, but not on an \EntityR{Investment Wallet}.
These validations follow the restrictions specified in
\cref{subsec:requirement_transaction_lifecycle_management}.

    
\subsubsection*{Transaction status} 
A \EntityR{Transaction} is created with the \Status{Processing}, indicating
they have not yet gone through validation or execution.
The \Status{Failed} is a terminal state,
representing a permanent failure with no possibility of retry.
Similarly, the \Status{Completed} is a terminal state,
representing that all validations and execution steps succeeded,
the transaction has settled, and the involved \Entity{Wallet} was updated.
The \Status{Transient Error} is a cyclic state, representing a recoverable
error that occurred during transaction processing (such as a failure in a
third-party API call), which is eligible for retry.
            
\subsection{Journal Entry}
\label{subsec:entity_journal_entry}

A \Entity{Journal\;Entry} documents the effects of a \EntityR{Transaction}
on a \EntityR{Wallet}. It tracks changes in the \EntityR{Wallet} 
\Variable{balance}.
By listing all \Entities{Journal\;Entry} linked to a
\EntityR{Wallet}, it is possible to calculate its current state
accurately.
    
\Entities{Journal\;Entry} are always created in pairs to represent
the source and destination of funds. To accurately reflect changes in a
\EntityR{Wallet}, a \EntityR{Journal\;Entry} includes the following:

\begin{itemize}
  \item \texttt{wallet\_id}:
        It identifies the source or target \EntityR{Wallet}.
        It may be absent to represent an external bank account.

  \item \texttt{subwallet\_id}:
        It identifies the \EntityR{Subwallet} under a \textsc{Wallet}.
        It may also be absent to represent an external bank account.

  \item \texttt{amount}:
        It accounts for the money being added (if positive) or deducted
        (if negative) from the \EntityR{Subwallet}.

  \item \texttt{balance\_type}:
        It can have three values:
        \TypeR{Internal}, for funds sent to or received from
        an external bank account;
        \TypeR{Available}, for funds added to or deducted from
        the available balance of a \EntityR{Subwallet}; and
        \TypeR{Holding}, for funds added to or deducted from
        the reserved balance of a \EntityR{Subwallet}.

\end{itemize}

\begin{figure}[b!]
  \centering
  \includegraphics[width=0.81\linewidth]{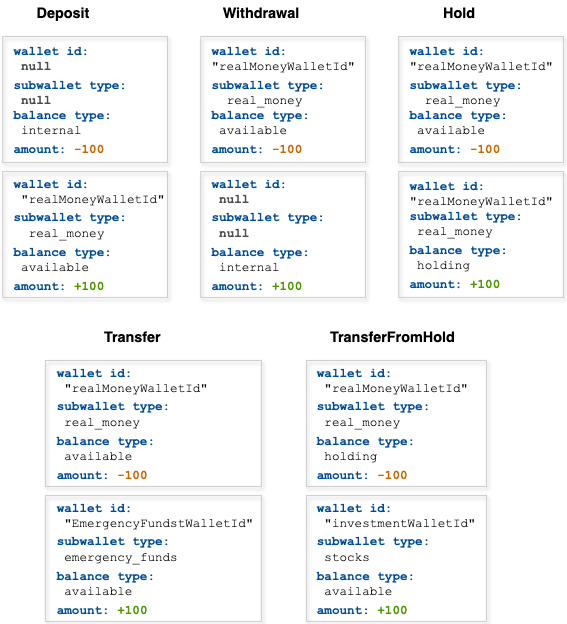}
  \Description{
    Pairs of \EntityR{Journal\;Entry} created for each
    \EntityR{Transaction} type.
  }
  \caption{
    Pairs of \EntityR{Journal\;Entry} for each
    \EntityR{Transaction} type.
  }
  \label{fig:journal_entries_per_transaction_type}
\end{figure}

The \Entity{Ledger} collects all \Entity{Journal\;Entry} created by the
\DW system. Each \EntityR{Transaction} with \Status{Completed} results
in a pair of \EntityR{Journal Entry} being recorded in the \EntityR{Ledger}.
\Cref{fig:journal_entries_per_transaction_type} illustrates how these
journal entries look like for each transaction type.

\subsection{Investment Policy}
\label{subsec:entity_investment_policy}

An \Entity{Investment\;Policy} defines the allocation of an investment portfolio
by associating each \EntityR{Subwallet} with a specific percentage representing
its share of the total portfolio, as specified in
\cref{subsec:requirement_investment_customization}.

Only a \EntityR{Subwallet} within the \EntityR{Investment Wallet} is
permitted to be included in an \EntityR{Investment Policy}.

\section{\DW System: Services}
\label{sec:system_services}

The \DW system relies on a set of services that collectively operate the
functionalities described in \cref{sec:requirements}.

\begin{figure}[t]
  \centering
  \includegraphics[width=0.90\linewidth]{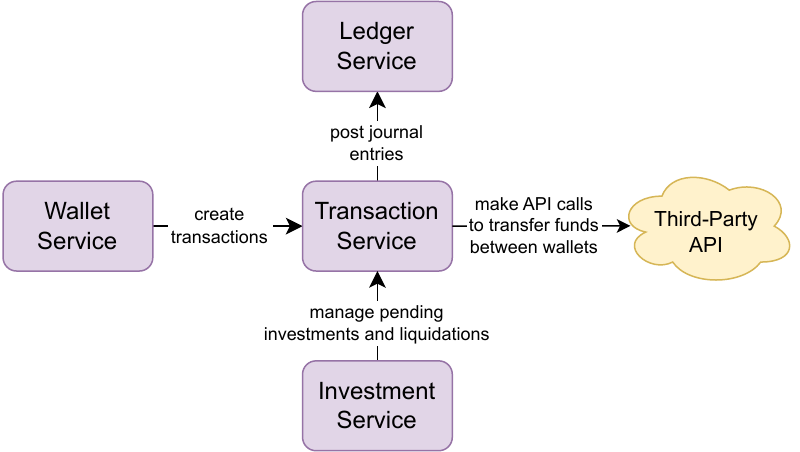}
  \Description{
    Services that compose the \DW system.
  }
  \caption{
    Services that compose the \DW system.
  }
  \label{fig:system_services}
\end{figure}

\Cref{fig:system_services} summarizes the existing services and illustrates
the communication flow between them.

\subsection{Ledger Service}
\label{subsec:service_ledger}

The \Service{Ledger} is responsible for orchestrating between other services
and the \DW system \Entity{Ledger}. It provides two functionalities:
  (1) posting new \EntityR{Journal\;Entry}, and
  (2) querying the \EntityR{Ledger} for its \Variable{balance}.
For the latter, it aggregates \EntityR{Journal\;Entry} to compute the balance
of a specific \EntityR{Wallet}.

\subsection{Investment Service}
\label{subsec:service_investment}

The \Service{Investment} is dedicated to handling investment and liquidation
requests, which involve multiple steps and do not settle instantly like other
transaction types. This service knows how to effectively use an
\EntityR{Investment\;Policy} to initiate and manage the transactions required to
complete an investment or liquidation.

\subsection{Transaction Service}
\label{subsec:service_transaction}

The \Service{Transactions} provides functionalities to fulfill each step
of a transaction lifecycle
specified in \cref{subsec:requirement_transaction_lifecycle_management}.
It creates, validates, and processes \EntityR{Transactions}, encapsulating
the business logic for each \EntityR{Transaction} type.

This service also guarantees that the \EntityR{Ledger} is accurately updated
upon \EntityR{Transaction} completion, maintaining consistency across the
system.	
    
Although the \Service{Transactions} is responsible for managing different
\EntityR{Transaction} types, it remains agnostic to their specific purposes.
Its role is to ensure that all transactions are processed correctly and that
the ledger remains consistent.

\subsection{Wallet Service}
\label{subsec:service_wallet}
    
The \Service{Wallets} is responsible for managing and initiating
\EntityR{Transaction}s within \EntityR{Wallet}s.
It encapsulates the business logic needed to translate money movements
initiated by the customer into a series of \EntityR{Transaction}s that
fulfill these requests.

The types of requests handled by the \Service{Wallets} include:

\begin{itemize}
  \setlength{\itemsep}{0.5\baselineskip}
  \item \Request{Deposit}:
        Adds funds into the \EntityR{Real Money Wallet}.
  \item \Request{Withdraw}:
        Removes funds from the \EntityR{Real Money Wallet}.
  \item \Request{Emergency Allocation}:
        Adds funds into the \EntityR{Emergency Funds Wallet}.
  \item \Request{Emergency Release}:
        Transfers funds from the \EntityR{Emergency Funds Wallet}
        to the \EntityR{Real Money Wallet}.
  \item \Request{Investment}:
        Transfers funds from the \EntityR{Real Money Wallet} to the
        \EntityR{Investment Wallet}, distributing the amount across investment
        options based on the \EntityR{InvestmentPolicy}.
  \item \Request{Liquidation}:
        Transfers funds from the \EntityR{Investment Wallet} to the
        \EntityR{Real Money Wallet}, liquidating investments proportionally
        according to the \EntityR{InvestmentPolicy}.
\end{itemize}

\section{Qualitative Analysis}\label{sec:qualitative_analysis}

The qualitative analysis compares functionally equivalent code
snippets from the Kotlin and Scala proofs of concept, evaluating
them according to the architectural characteristics focused on by
the research sub-questions. This methodology is self-ethnographic,
since it presents the assessments of the authors who led the
development process.

\subsection{Extensibility [\sRQ{1}]}
\label{subsec:qualitative_analysis_extensibility}

To analyze the \emph{Extensibility} quality attribute, this section examines
the Kotlin and Scala implementations of a \Class{Wallet} in the \DW system.

\subsubsection*{Kotlin}

\begin{code}[t]
  \InputKotlin[
    label=( Kotlin --- Wallet ),
  ]{code/kotlin/Wallet.kt}
  \InputKotlin[
    label=( Kotlin --- Real Money Wallet ),
  ]{code/kotlin/RealMoneyWallet.kt}
  \InputKotlin[
    label=( Kotlin --- Emergency Funds Wallet ),
  ]{code/kotlin/EmergencyFundsWallet.kt}
  \InputKotlin[
    label=( Kotlin --- Investment Wallet ),
  ]{code/kotlin/InvestmentWallet.kt}
  \vspace{-\baselineskip}
  \Description{
    \Class{Wallet} class \emph{inheritance} hierarchy in Kotlin
  }
  \caption{
    \Class{Wallet} class \emph{inheritance} hierarchy in Kotlin
  }
  \label{lst:kt:wallet_hierarchy}
\end{code}

In Kotlin, the \Entity{Wallet} is structured using OOP principles like
inheritance and polymorphism. The \emph{abstract class} \Class{Wallet}
serves as a blueprint for all specific wallet types: \Class{Real Money Wallet},
\Class{Investment Wallet}, and \Class{Emergency Funds Wallet}.
This is illustrated in \cref{lst:kt:wallet_hierarchy}.

In Kotlin, \Method{getAvailableBalance} is an abstract method of the
\Class{Wallet} class, which encapsulates the logic for querying balances from
the ledger. All subclasses must implement their specific logic for computing
balances. This ensures consistency across wallet types while allowing each
subclass to customize its behavior. Such an approach leverages polymorphism:
the behavior of a method depends on the subclass of the object being called.

A future requirement demanding a new \Class{Wallet} could be achieved by
simply adding a new subclass that inherits from the \Class{Wallet} abstract
class. This subclass automatically gains access to the shared attributes
defined in the superclass while also being required to implement its
abstract method, enabling it to define specific balance-query behavior. 

\subsubsection*{Scala}
In Scala, the \Entity{Wallet} is structured using FP principles like
immutable data structures. The \emph{case class} \Class{Wallet} consolidates
all possible attributes and contains an additional field \Object{WalletType}
that distinguishes the wallet derivations within the system.
This is illustrated in \cref{lst:sc:wallet_and_wallet_type}.

\begin{code}[t]
  \InputScala[
    numbers=left,
    label=( Scala --- Wallet ),
  ]{code/scala/Wallet.sc}
  \InputScala[
    numbers=left,
    label=( Scala --- WalletType ),
  ]{code/scala/WalletType.sc}
  \InputScala[
    numbers=left,
    label=( Scala --- getAvailableBalance ),
  ]{code/scala/getAvailableBalance.sc}
  \vspace{-\baselineskip}
  \Description{
    \Class{Wallet} type and factory \emph{pure function} in Scala
  }
  \caption{
    \Class{Wallet} type and factory \emph{pure function} in Scala
  }
  \label{lst:sc:wallet_and_wallet_type}
  \vspace{-\baselineskip}
\end{code}

In Scala, \texttt{getAvailableBalance} is a standalone function in the
\Class{WalletService} class, which encapsulates the logic for querying balances
from the ledger.  For every variant of the \texttt{WalletType} enumeration, the
function specifies a corresponding list of ledger queries to compute the
available balance. This provides behavior tailored to each wallet type without
requiring separate case classes. Such an approach leverages pattern matching:
the \texttt{walletType} field maps to different balance computation logic.

A future requirement demanding a new \Class{Wallet} could be achieved by simply
adding a new \Class{WalletType} enumeration. In Scala, pattern matching is
exhaustive, meaning that any unhandled enumeration variant results in a
compilation error. Therefore, pattern matching ensures that all standalone
functions related to \Class{WalletType} need to be revisited, ensuring it
to define all balance-query behavior.

\subsubsection*{Kotlin vs. Scala}
Both Kotlin and Scala implementations promote extensibility, enabling the system
to adapt through localized code updates. In Kotlin, this is achieved by adding a
new subclass to represent a wallet type and implementing the specialized
\Method{getAvailableBalance} method for that class. In Scala, extensibility is
achieved by introducing a new variant to the \Class{WalletType} enumeration.

These features also enforce the extension of existing specialized logic
whenever a new wallet type is introduced.
  While Kotlin requires the creation of a dedicated class and method
  implementation for each new type,
  in Scala this can be achieved by simply adding a new enumeration variant
  and updating the existing functions that centralize the behavior
  of all wallet types.

\subsection{Reusability [\sRQ{2}]}
\label{subsec:qualitative_analysis_reusability}

To analyze the \emph{Reusability} quality attribute, this section examines
the Kotlin and Scala implementations of retrying a batch of transactions.
Not all transaction failures within the \DW system are permanent. For example,
transactions that pass validation but fail during execution are eligible
for retry.

The \Method{retryBatch} method attempts to reprocess each transaction within a
batch up to \texttt{n} times. If any transaction fails while interacting with
the partner API, the entire \Method{retryBatch} operation should fail.
Conversely, if all transactions are successfully processed, the batch
should be marked as successful.

\subsubsection*{Kotlin}

\begin{code}[b]
  \InputKotlin[
    numbers=left,
    label=( Kotlin --- retryBatch ),
  ]{code/kotlin/retryBatch.kt}
  \vspace{-\baselineskip}
  \caption{\texttt{retryBatch} built with a \texttt{for} loop in Kotlin}
  \label{lst:kt:retryBatch}
\end{code}

In Kotlin, the implementation of the retry logic is built around a
\texttt{while} loop. This is illustrated in \cref{lst:kt:retryBatch}.
It leverages the counter variable \texttt{attempts} to track the number
of retries for each transaction and the flag variable \texttt{success}
to indicate whether the transaction processing succeeded.

This design lacks the flexibility to be reused for other operations,
even though it provides the necessary control for transactions.

\subsubsection*{Scala}

\begin{code}[t]
  \InputScala[
    numbers=left,
    label=( Scala --- retryBatch ),
  ]{code/scala/retryBatch.sc}
  \InputScala[
    numbers=left,
    label=( Scala --- retry ),
  ]{code/scala/retry.sc}
  \vspace{-\baselineskip}
  \caption{\texttt{retryBatch} built over the HOF \texttt{retry} in Scala}
  \label{lst:sc:retryBatch}
\end{code}

In Scala, the implementation of the \texttt{retry} logic is built around
a higher-order function named \texttt{retry}. This is illustrated in
\cref{lst:sc:retryBatch}. It takes a function \texttt{f:\;=>\;Either[A,\;B]}
and an integer \texttt{n} as parameters to execute \texttt{f} up to
\texttt{n} times, early returning  \texttt{Right} if any attempt succeeds,
or returning \texttt{Left} if all \texttt{n} attempts fail.

This design encapsulates the retry logic into a reusable function,
providing this functionality to any function that needs to be retried.

\subsubsection*{Kotlin vs. Scala}

This example demonstrates how the concept of higher-order functions enables
the reuse of the same logic across multiple operations in the Scala PoC, 
eliminating the need to replicate similar logic for different functions.
By leveraging pattern matching and recursion, the Scala PoC also achieves
a more concise and expressive code structure compared to the Kotlin PoC.

\subsection{Error Handling and Propagation [\sRQ{3}]}
\label{subsec:qualitative_analysis_error_management}

To analyze error handling and propagation, this section examines how Kotlin
and Scala implement the logic to create investments. 

An investment allocates funds from the \Entity{Real Money Wallet} to the
\Entity{Investment Wallet}. This operation consists essentially of a hold
\EntityR{Transaction} that does not settle instantly.  

\subsubsection*{Kotlin}

\begin{code}[b]
  \vspace{-1em}
  \InputKotlin[
    numbers=left,
    label=( Kotlin --- invest ),
  ]{code/kotlin/invest.kt}
  \vspace{-\baselineskip}
  \caption{\texttt{invest} method using \textit{exceptions} in Kotlin}
  \label{lst:kt:invest}
\end{code}

In Kotlin, the hold \Transaction{Transaction} is created through the
\Method{processTransaction} method, which can throw exceptions.
These exceptions must be explicitly caught in the \Method{invest}
method to map them to a corresponding \Class{WalletService} exception.
Additionally, within the \texttt{catch} blocks, it is necessary to ensure
the \Transaction{Transaction} transitions to the appropriate state.
This is illustrated in \cref{lst:kt:invest}.

However, if a new exception type is created in the future,
developers must explicitly update the \texttt{try-catch} block,
adding a new catch case, otherwise risking unhandled runtime
exceptions if they forget about it.

Additionally, as a \texttt{try-catch} block might not handle
all the error cases, reasoning about the program becomes more challenging
since the code does not clearly express the potential failure paths.

\subsubsection*{Scala}

\begin{code}[t]
  \InputScala[
    numbers=left,
    label=( Scala --- invest ),
  ]{code/scala/invest.sc}
  \vspace{-\baselineskip}
  \caption{\texttt{invest} function using \textit{monadic types} in Scala}
  \label{lst:sc:invest}
\end{code}

In Scala, errors are emitted via the monadic type \texttt{Either}.
This approach allows them to be explicitly modeled as part of the method's
return type, making it clear to developers that a method can fail.
This is illustrated in \cref{lst:sc:invest}.

The use of \texttt{Either} enhances type safety by enforcing that the caller
must handle both success and failure cases, ensuring that no error is missed
during development. This promotes more predictable and reliable error handling
within the system.

\section{Quantitative Analysis}
\label{sec:quantitative_analysis}

The quantitative analysis aims to gather feedback from developers with diverse
backgrounds through a survey, complementing the qualitative analysis. 

The draft survey was initially reviewed by a developer with experience in both
paradigms. Then, it was beta-tested with two additional developers knowledgeable
in both paradigms. After incorporating all feedback, the final version was
made publicly available to a wider public.

The survey was divided into two sections:

\subsubsection*{Background}
This section was designed to map the developers' experience with object-oriented
and functional programming languages, since such experience might influence
participants' reasoning when evaluating the code snippets.

\subsubsection*{Code Analysis}
This section was designed to measure the developer's impressions about the
two \DW system proof of concepts. Each question includes a brief explanation
of a functionality from the \DW system, followed by two code snippets:
  the first from the Kotlin PoC and
  the second from the Scala PoC.
For each code snippet, participants were asked to evaluate
the six architectural characteristics summarized in
\cref{tab:architectural_characteristics}. They answered the question:
\emph{``From the code, how easy is it to...''} using a Likert scale with
five steps, from \emph{Very Easy} to \emph{Very Hard}.

\medskip
At the publication of the bachelor thesis that inspired this paper%
~\cite{Dias2024FunctionalSystems}, the survey received \emph{eight responses}.
Considering space constraints, this section will only address two of four
questions proposed in the questionnaire, which intersect with some of
the code previously analyzed in \cref{sec:qualitative_analysis}.
For a full analysis, please refer to the bachelor thesis%
~\cite{Dias2024FunctionalSystems}.

The full questionnaire and corresponding responses can be found in the
paper's \href{https://doi.org/10.5281/zenodo.16618132}{reproduction
package}.

\subsection{Question \#1 --- \Method{retryBatch}}
\label{subsec:quantitative_analysis_retryBatch}

\begin{figure}[b!]
  \centering
  \includegraphics[width=0.95\linewidth]{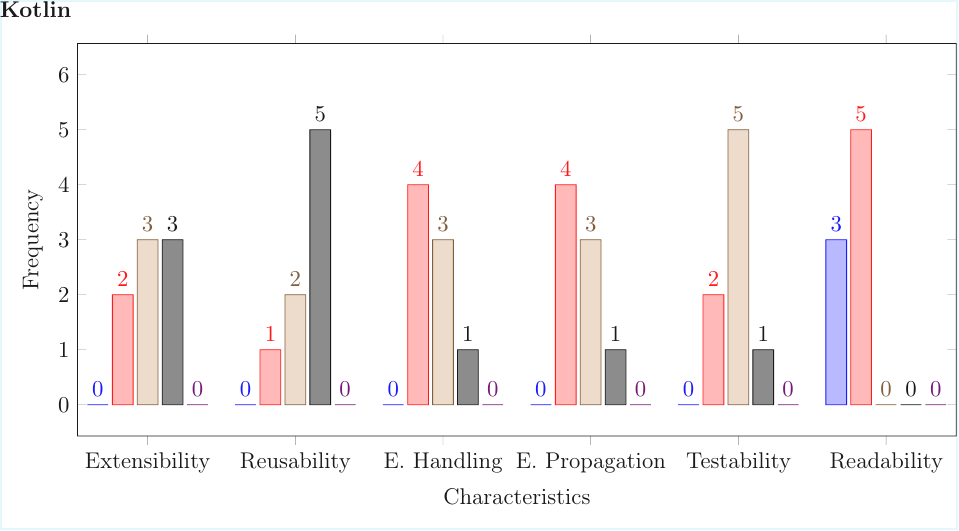}
  \includegraphics[width=0.95\linewidth]{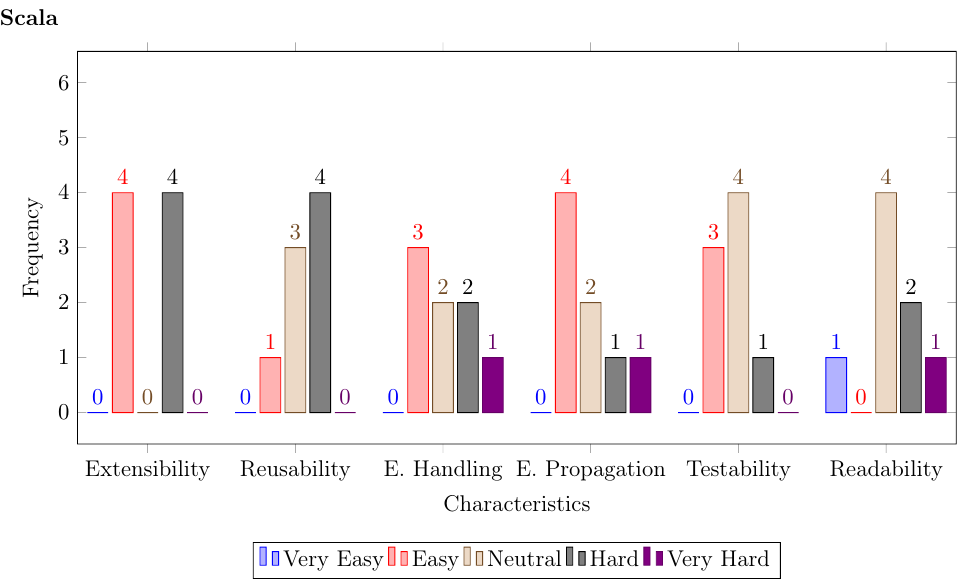}
  \Description{Kotlin vs. Scala --- \Method{retryBatch}}
  \caption{Kotlin vs. Scala --- \Method{retryBatch}}
  \label{g:q1_answers}
\end{figure}

\hyperref[g:q1_answers]{Question \#1} collected feedback on the implementation
of the \Method{retryBatch}, whose code was explored in
\cref{lst:kt:retryBatch,lst:sc:retryBatch}.

The results of \hyperref[g:q1_answers]{Question \#1} are shown in
\cref{g:q1_answers}. They reveal a notable difference in the architectural
characteristics of \emph{error handling}, \emph{error propagation}, and
\emph{readability}.

Many developers found the Scala implementation of the \texttt{retry} logic
to be more challenging to handle, to propagate errors, and to read when
compared to the Kotlin implementation. This indicates that,
for these aspects, Kotlin may offer a clearer approach.

\subsection{Question \#3 --- \Method{invest}}
\label{subsec:quantitative_analysis_invest}

\begin{figure}[t!]
  \centering
  \includegraphics[width=0.95\linewidth]{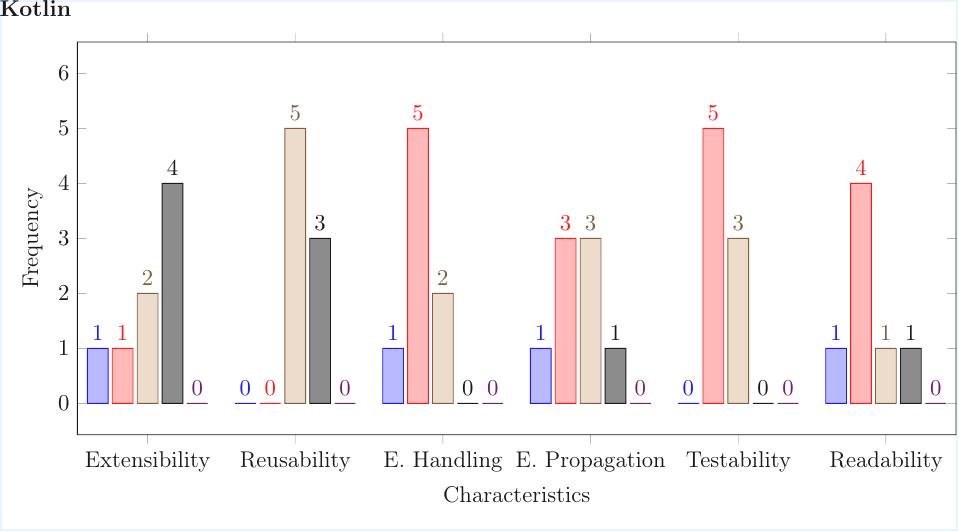}
  \includegraphics[width=0.95\linewidth]{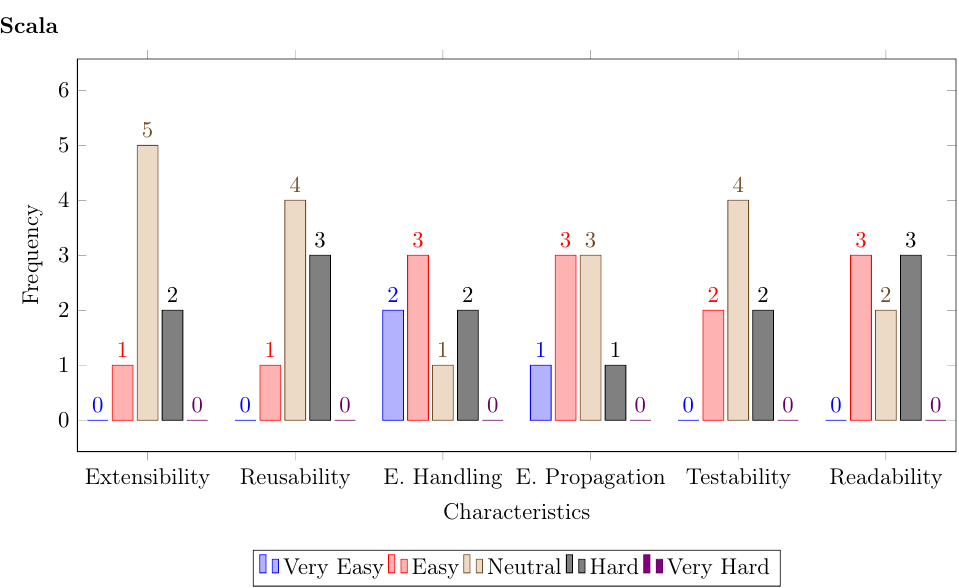}
  \Description{Kotlin vs. Scala --- \Method{invest}}
  \caption{Kotlin vs. Scala --- \Method{invest}}
  \vspace{-1.3em}
  \label{g:q3_answers}
\end{figure}

\hyperref[g:q3_answers]{Question \#3} collected feedback on the implementation
of the \Method{invest} logic, whose code was explored in
\cref{lst:kt:invest,lst:sc:invest}.

The results of the \hyperref[g:q3_answers]{Question \#3} are shown
in \cref{g:q3_answers}. They reveal that \emph{extensibility} is slightly
favored in the Scala PoC compared to the Kotlin PoC. On the other hand,
\emph{testability} showed a significant advantage in favor of Kotlin.

\emph{Error handling} and \emph{error propagation} were comparable
between both implementations, though some responses suggested these
aspects might be more challenging to achieve in Scala.
\emph{Reusability} was also quite similar for both implementations,
with a slight preference for Scala.  \emph{Readability}, while
inconclusive for Scala, received more favorable feedback for Kotlin.

\section{Discussion}
\label{sec:discussion}

Based on the findings from the qualitative and quantitative analyses,
this section discusses the research subquestions.



\begin{RevisitSubResearchQuestion}{1}
  How do the functional and object-oriented paradigms impact
  the extensibility of a system? 
\end{RevisitSubResearchQuestion}

The qualitative analysis revealed that both Kotlin and Scala offer features
that promote extensibility. Kotlin leveraged object-oriented principles such
as inheritance and polymorphism to achieve extensibility, while Scala employed
enumerations and pattern matching to accomplish the same objective. Although the
Scala implementation was more concise, requiring less boilerplate code to extend
a model, both approaches were effective and provided comparable benefits to the
extensibility characteristic.

The quantitative analysis corroborated these findings. The perceptions of Scala
and Kotlin implementations were similar, with only minor differences observed.
Notably, these differences mostly favored the Kotlin implementation of the
Digital Wallet System.

\begin{RevisitSubResearchQuestion}{2}
How do the functional and object-oriented paradigms impact
the reusability of a system? 
\end{RevisitSubResearchQuestion}

Scala demonstrated stronger support for code reusability through higher-order
functions, enabling generic implementations of various functionalities. In
contrast, the Kotlin implementations in both examples lacked sufficient
flexibility to be effectively adapted for different domains.

However, the quantitative analysis did not reveal significant differences
between Kotlin and Scala in terms of reusability. Responses indicated that both
implementations were perceived as similar, with only minor variations.

The discrepancy between the qualitative and quantitative analysis findings
suggests that the reusability characteristic may considerably depend on the
specific functionality being evaluated.

\begin{RevisitSubResearchQuestion}{3}
  How do the functional and object-oriented paradigms impact
  error handling and propagation of a system? 
\end{RevisitSubResearchQuestion}

Kotlin propagates errors using exceptions and handles them with traditional
\texttt{try-catch} blocks. In contrast, Scala utilizes the monadic type
\texttt{Either[A, B]} to propagate errors. Error handling in Scala is performed
through higher-order functions, allowing mapping to new error types as needed.

By incorporating errors directly into return types, the Scala implementation
enforces comprehensive error handling through type checking, ensuring that new
error cases cannot be missed.

Nevertheless, the quantitative analysis did not reflect the same dominance of
the Scala implementation. Most responses indicated that Kotlin is either
comparable or slightly superior to Scala in terms of error handling and
propagation.


\section{Threats to Validity}\label{sec:threats_to_validity}


The threats to validity acknowledge the limitations of this research.

\subsubsection*{Construct Validity}\label{subsec:construct_validity}
The evaluation of programming paradigms in this research may be influenced by
the choice of programming languages representing each paradigm. Kotlin and Scala
were chosen as modern, multi-platform languages that run on the Java Virtual
Machine (JVM), ensuring a level of syntactic compatibility between the
programming languages. Despite Kotlin supporting FP features and Scala
supporting OOP features, this research intentionally limited both to
their selected domains.

\subsubsection*{External Validity}\label{subsec:external_validity}
The qualitative analysis conducted in this research may reflect biases
influenced by the author's background and expertise. To address this, a
mixed-methods approach was employed, incorporating quantitative analysis to
complement the qualitative findings. Nevertheless, due to time constraints,
the number of participants in the survey was limited. To strengthen the
results, more participants should be included to triangulate the
conclusions from this research.

\section{Conclusion}
\label{sec:conclusion}

\begin{RevisitMainResearchQuestion}
  How do the functional and object-oriented paradigms impact different
  architectural characteristics of a system?
\end{RevisitMainResearchQuestion}

This paper compared the impact of adopting OOP versus FP over six architectural
characteristics of software systems. It introduced the \DW system, implemented
twice: once in Kotlin (OOP), once in Scala (FP). The use of these modern,
popular programming languages for the research will hopefully reach the
interest of practitioners, who may understand the results and then be
better informed when choosing a certain paradigm.


For future work, this research can be extended to target other architectural 
characteristics, such as scalability and security. Additionally, the current
results can be enhanced by expanding the survey, thus addressing some
of the threats to validity.


\bibliographystyle{ACM-Reference-Format}
\bibliography{main}

\end{document}